\pdfoutput=1

\documentclass[acmsmall,screen,10pt]{acmart}
\usepackage{amsmath,xcolor,cleveref, bm}
\usepackage{xspace}
\usepackage{siunitx,wrapfig}

\renewcommand{\>}{\quad\quad}
\newcommand{\K}[1]{\ensuremath{\mathbf{#1}}}
\newcommand{\F}[1]{\ensuremath{\mathsf{#1}}}
\newcommand{\MB}{\ensuremath{\,\text{MB}}}

\setcopyright{rightsretained}
\acmJournal{PACMPL}
\acmYear{2022}
\acmVolume{6}
\acmNumber{OOPSLA2}
\acmArticle{160}
\acmMonth{10}
\acmPrice{}
\acmDOI{10.1145/3563323}
\copyrightyear{2022}
\acmSubmissionID{oopslab22main-p349-p}

\begin{document}

\title{Optimal Heap Limits for Reducing Browser Memory Use}

\author{Marisa Kirisame}
\affiliation{
  \department{Computer Science}
  \institution{University of Utah}
  \country{United States}
}
\email{marisa@cs.utah.edu}          

\author{Pranav Shenoy}
\affiliation{
  \department{Computer Science}
  \institution{University of Utah}
  \country{United States}
}
\email{pranav.shenoy@utah.edu}         

\author{Pavel Panchekha}
\affiliation{
  \department{Computer Science}
  \institution{University of Utah}
  \country{United States}
}
\email{pavpan@cs.utah.edu}         

\def\nameA{splay.js\xspace}
\def\lA{\num{31}\xspace}
\def\gA{\num{633}\xspace}
\def\sA{\num{525}\xspace}
\def\mbextraA{\num{50}\xspace}
\def\baseextraA{\textbf{30}\xspace}
\def\mbruntimeA{\textbf{32}\xspace}
\def\baseruntimeA{\num{36}\xspace}
\def\mbgctimeA{\textbf{51}\xspace}
\def\basegctimeA{\num{77}\xspace}
\def\nameB{typescript.js\xspace}
\def\lB{\num{30}\xspace}
\def\gB{\num{57}\xspace}
\def\sB{\num{440}\xspace}
\def\mbextraB{\textbf{21}\xspace}
\def\baseextraB{\num{28}\xspace}
\def\mbruntimeB{\textbf{30}\xspace}
\def\baseruntimeB{\num{30}\xspace}
\def\mbgctimeB{\num{5.9}\xspace}
\def\basegctimeB{\textbf{4.0}\xspace}
\def\nameC{pdfjs.js\xspace}
\def\lC{\num{96}\xspace}
\def\gC{\num{34}\xspace}
\def\sC{\num{383}\xspace}
\def\mbextraC{\textbf{22}\xspace}
\def\baseextraC{\num{79}\xspace}
\def\mbruntimeC{\num{33}\xspace}
\def\baseruntimeC{\textbf{33}\xspace}
\def\mbgctimeC{\num{11}\xspace}
\def\basegctimeC{\textbf{4.6}\xspace}
\def\nameD{Total\xspace}
\def\lD{\num{158}\xspace}
\def\gD{\num{725}\xspace}
\def\sD{\num{1350}\xspace}
\def\mbextraD{\textbf{95}\xspace}
\def\baseextraD{\num{138}\xspace}
\def\mbruntimeD{\textbf{96}\xspace}
\def\baseruntimeD{\num{100}\xspace}
\def\mbgctimeD{\textbf{68}\xspace}
\def\basegctimeD{\num{85}\xspace}
\def\JSSplayPDFJSl{\num{0.3}\xspace}
\def\JSSplayPDFJSg{\num{18}\xspace}
\def\JSSplayPDFJSs{\num{1.4}\xspace}
\def\JSSplayPDFJSGCFreq{\num{8.9}\xspace}
\def\JSSplayPDFJSExtraMemSquared{\num{4.3}\xspace}
\def\JSSplayPDFJSExtraMem{\num{2.1}\xspace}

\def\SingleEvalSCROLLPIX{\num{50}\xspace}
\def\SingleEvalSCROLLSLEEP{\num{1}\xspace}
\def\SingleEvalEVALSLEEP{\num{5}\xspace}
\def\SingleEvalGMAILWAITTIME{\num{5}\xspace}
\def\SingleEvalGMAILINBOXTIME{\num{10}\xspace}
\def\SingleEvalGMAILEMAILTIME{\num{5}\xspace}

\def\JSMinC{\SI{3.0}{\%/MB}\xspace}
\def\JSMaxC{\SI{30.0}{\%/MB}\xspace}
\def\WEBMinC{\SI{0.05}{\%/MB}\xspace}
\def\WEBMaxC{\SI{0.9}{\%/MB}\xspace}
\def\ACDCMinC{\SI{0.1}{\%/MB}\xspace}
\def\ACDCMaxC{\SI{10.0}{\%/MB}\xspace}

\def\BiasInWorkingMemoryInMB{5\xspace}
\def\ExtraFloorInMB{3\xspace}
\def\GCSpeedSmoothingPerSample{0.5\xspace}
\def\GarbageRateDecayPerSec{0.95\xspace}
\def\InitialGCSpeed{100.0\xspace}
\def\MinHeapExtraSizeInMB{2\xspace}
\def\TotalMemoryFloorInMB{10\xspace}

\def\MBHash{cdf33a3ad4992aba6ba8c7e546b1169df12341d6\xspace}
\def\VEightHash{159b6c337fb651417910c22b4ad43b76a835c707\xspace}
\def\JSMaxSpeedup{\num{140.0}\%\xspace}
\def\JSMaxSaving{\num{7.0}\%\xspace}
\def\CompareAt{\num{20.0}\xspace}
\def\JETSTREAMSpeedup{\num{37.0}\%\xspace}
\def\JETSTREAMMemorySaving{\num{8.8}\%\xspace}
\def\JETSTREAMImprovement{\num{3.4}\,\sigma\xspace}
\def\JETSTREAMPValue{\num{9.9e-07}\xspace}
\def\JETSTREAMMaxRegress{\num{-2.1}\,\sigma\xspace}
\def\JETSTREAMMaxImprovement{\num{4.9}\,\sigma\xspace}
\def\BROWSERIIISpeedup{\num{48.0}\%\xspace}
\def\BROWSERIIIMemorySaving{\num{16.0}\%\xspace}
\def\BROWSERIIIImprovement{\num{2.2}\,\sigma\xspace}
\def\BROWSERIIIPValue{\num{7.7e-81}\xspace}
\def\BROWSERIIIMaxRegress{\num{0.068}\,\sigma\xspace}
\def\BROWSERIIIMaxImprovement{\num{5.7}\,\sigma\xspace}
\def\ACDCSpeedup{\num{30.0}\%\xspace}
\def\ACDCMemorySaving{\num{8.2}\%\xspace}
\def\ACDCImprovement{\num{3.3}\,\sigma\xspace}
\def\ACDCPValue{\num{3.3e-12}\xspace}
\def\ACDCMaxRegress{\num{-1.2}\,\sigma\xspace}
\def\ACDCMaxImprovement{\num{4.9}\,\sigma\xspace}
\def\BROWSERISpeedup{\num{67.0}\%\xspace}
\def\BROWSERIMemorySaving{\num{11.0}\%\xspace}
\def\BROWSERIImprovement{\num{2.0}\,\sigma\xspace}
\def\BROWSERIPValue{\num{1.6e-16}\xspace}
\def\BROWSERIMaxRegress{\num{-0.066}\,\sigma\xspace}
\def\BROWSERIMaxImprovement{\num{5.0}\,\sigma\xspace}
\def\BROWSERIISpeedup{\num{49.0}\%\xspace}
\def\BROWSERIIMemorySaving{\num{16.0}\%\xspace}
\def\BROWSERIIImprovement{\num{1.4}\,\sigma\xspace}
\def\BROWSERIIPValue{\num{1.1e-52}\xspace}
\def\BROWSERIIMaxRegress{\num{0.68}\,\sigma\xspace}
\def\BROWSERIIMaxImprovement{\num{5.5}\,\sigma\xspace}
\def\GraphHash{1cd0382a970fb4ab2080540a5201cdc1f169de0c\xspace}

\begin{abstract}
Garbage-collected language runtimes
  carefully tune heap limits
  to reduce garbage collection time
  and memory usage.
However, there's a trade-off:
  a lower heap limit reduces memory use
  but increases garbage collection time.
Classic methods for setting heap limits
  include manually tuned heap limits
  and multiple-of-live-size rules of thumb,
  but it is not clear
  when one rule is better than another
  or how to compare them.

We address this problem with a new framework
  where heap limits are set
  for multiple heaps at once.
Our key insight is that every heap limit rule
  induces a particular allocation of memory
  across multiple processes,
  and this allocation can be sub-optimal.
We use our framework
  to derive an optimal
  ``square-root'' heap limit rule,
  which minimizes total memory usage
  for any amount of total garbage collection time.
Paradoxically, the square-root heap limit rule
  achieves this coordination without communication:
  it allocates memory optimally across multiple heaps
  without requiring any communication between those heaps.

To demonstrate that this heap limit rule is effective,
  we prototype it for V8, the JavaScript runtime
  used in Google Chrome, Microsoft Edge, and other browsers,
  as well as in server-side frameworks like node.js and Deno.
On real-world web pages,
  our prototype achieves reductions of approximately
  \BROWSERIIMemorySaving of memory usage
  while keeping garbage collection time constant.
On memory-intensive benchmarks,
  reductions of up to \ACDCSpeedup of garbage collection time
  are possible with no change in total memory usage.
  
\end{abstract}



\maketitle

\section{Introduction}

Many modern programming languages are garbage-collected~\cite{gcBook},
  freeing the programmer from manually managing memory.
Achieving good performance in a garbage-collected language,
  however,
  requires controlling how often garbage collection occurs
  and how long it takes.
Typically,
  garbage-collected language runtimes
  control garbage collection frequency
  by setting a limit on total memory usage (``heap size'')
  and collecting garbage once that limit is hit.%
\footnote{In production garbage-collected runtimes such as V8 or the JVM,
  there are lots of other triggers for garbage collection.
  Nevertheless, the heap limit does play a central role.}

Choosing a heap limit is a classic space-time trade-off.
Set the heap limit too low,
  and the garbage collector will fire too often,
  slowing the program unacceptably.
Set it too high,
  and the program consumes too much memory,
  interfering with other processes.
In high-performance deployments
  of garbage-collected systems,
  setting heap memory limits manually
  to achieve the optimal trade-off is common~\cite{Uber}.
However, in dynamic environments
  this manual tuning is untenable.
For example,
  web browsers start and stop applications frequently
  as their usually-inexpert users
  (i.e., your grandmother)
  browse the web;
  browsers must therefore
  set heap limits automatically,
  without relying on user expertise.

Heap limits affect memory usage,
  and memory usage is important to end users---%
  perhaps even to you, dear reader.
Memory usage is one of
  the most common complaints about web browsers~\cite{Reduce},
  with hundreds of articles instructing users
  on how to reduce their browser's memory usage,
  with suggestions like uninstalling extensions,
  disabling security features,
  and deleting old user data~\cite{Reduce}.
Major browser developers make significant efforts
  to reduce memory usage,
  such as Firefox's ``Are we Slim Yet'' effort~\cite{AWSY}
  and Chrome's ``V8 Lite'' effort~\cite{Lite}.
And because the JavaScript heap is typically
  a large percentage of overall browser memory usage,
  often more than half,
  choosing appropriate heap limits
  is central to lowering browser memory use.
This is all the more important
  on low-memory devices such as phones.
On these devices,
  lower heap limits can allow more tabs
  to stay resident in memory,
  which not only makes browsing more convenient,
  but is also necessary to complete
  multi-tab workflows such as some single sign-on systems.
Users of such devices cannot simply buy more memory,
  and even if they could, they may not have the money.

The challenge with deciding heap limits automatically
  is that it's unclear what to optimize for.
A typical approach is to set heap limits
  to some multiple of the live size~\cite{PerfGC}.
But since there is a trade-off
  between memory usage and program run time,
  it's not clear in what sense
  this rule of thumb is optimal or sub-optimal,
  or even how to compare it to some other approach.
That said,
  there's good reason to think
  that this rule is \emph{not} optimal.
State of the art garbage-collected language runtimes
  extend it with ad-hoc fixes,
  such as detecting when a program is idle
  and scheduling extra garbage collections~\cite{Idle}, capping/flooring the heap limit,
  or targeting a percentage of total run time.
These ad-hoc fixes
  themselves conflict in complex ways,
  such as a idle collections
  conflicting with a run time target
  or run time targets conflicting with heap limit caps.

\begin{figure}
    \centering
    \includegraphics[width=\linewidth,trim=.65in 4in 1.75in 1in,clip,page=3]{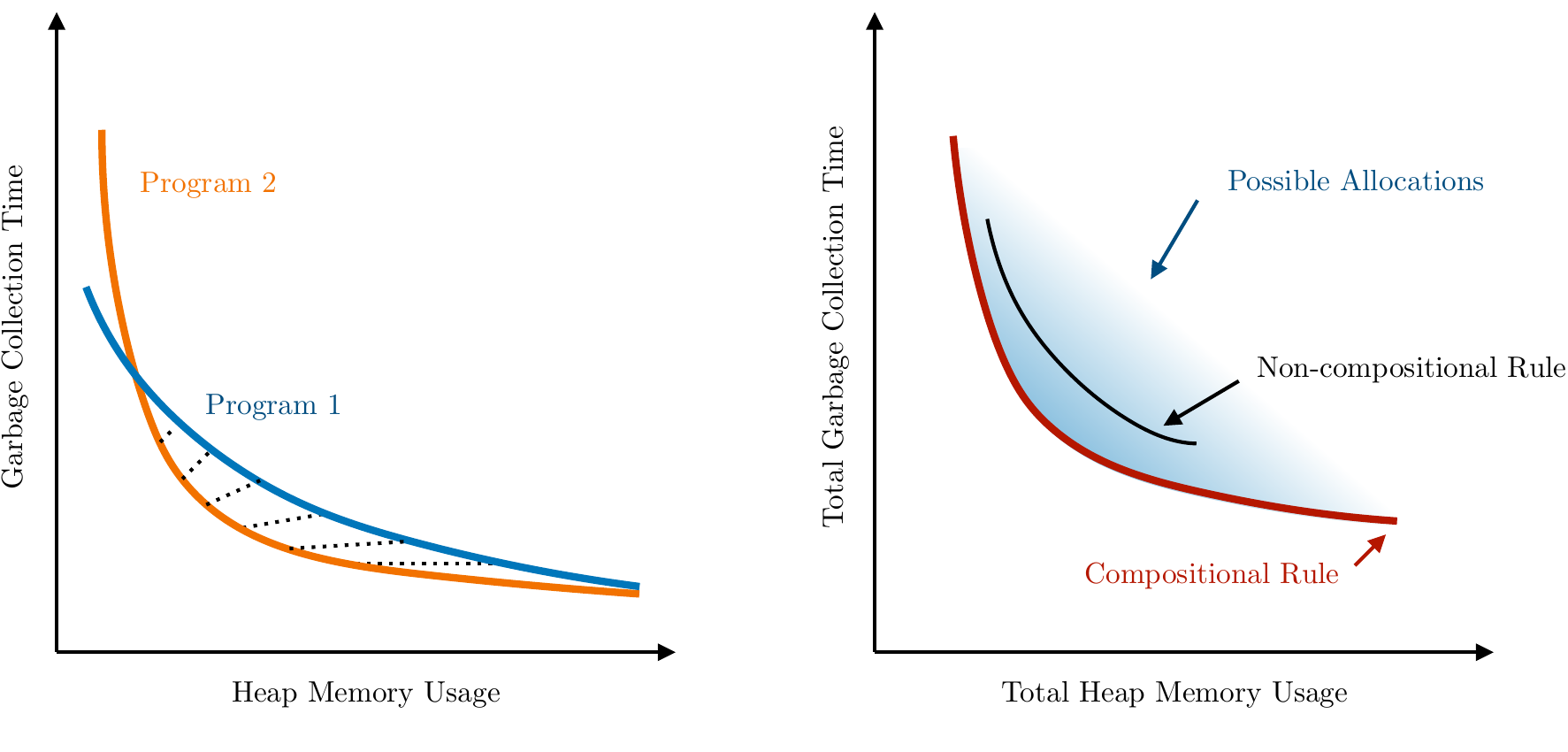}
    \caption{
    Synthetic data
      showing the difference
      between compositional
      and non-compositional
      heap limit rules.
    In the left-hand plot,
      the garbage collection time
      and heap memory usage
      for two different programs
      is shown for varying heap limits.
    In the right-hand plot,
      both programs are run simultaneously
      and total heap memory usage
      and total garbage collection time
      across both programs is measured.
    Depending on the heap limit chosen
      for each program,
      any point in the blue region
      is achievable.
    The dotted lines in the left-hand plot
      and the black curve in the right-hand plot
      represent the behavior
      of a non-compositional heap limit rule
      as some tuning parameter is varied.
    This heap limit rule is not at the pareto frontier
      (as shown on the right-hand plot)
      because non-compositional heap limit rules
      mis-allocate memory among the programs.
    This means that, no matter how the parameter is tuned,
      the optimal trade-off
      between heap memory usage and garbage collection time
      cannot be achieved with this heap limit rule.
    A compositional heap limit rule,
      by contrast,
      allocates memory across programs
      in a pareto-optimal way,
      shown by the red curve.
    }
    \label{fig:pareto}
\end{figure}

We introduce a new framework
  in which heap limit rules can be compared,
  based on the insight that heap limit rules
  not only affect the performance of a single program,
  but also the allocation of memory \emph{across} programs.
In our framework,
  a shared pool of memory must be split
  across multiple garbage-collected programs to minimize
  the total time spent in garbage collection.
Optimal heap limit rules for multiple heaps
  must allocate available memory amongst the heaps
  in a way that minimizes total garbage collection time.
We find that standard heap limit rules
  are \emph{not} optimal,
  no matter how they are tuned.
Therefore,
  running multiple garbage-collected programs
  using standard heap limit rules---%
  opening multiple tabs in a web browser, for example---%
  will use more memory and spend more time collecting garbage
  than necessary.
\Cref{fig:pareto} shows the process graphically:%
\footnote{
    Using synthetic data
    where garbage collection time
    is inversely proportional to heap memory usage,
    plus a horizontal/vertical offset.
}
  non-optimal heap limit rules
  misallocate memory between processes
  and thus do not achieve
  a pareto-optimal trade-off
  between memory usage
  and garbage collection time.

Luckily, our framework is analytically solvable,
  producing a ``square-root'' heap limit rule
  that \emph{is} optimal in our framework.
This means that multiple processes that each use
  identically-tuned square-root heap limit rules
  allocate memory amongst themselves
  so as to minimize total garbage collection time.
Surprisingly,
  this does not require any communication
  between the processes,
  achieving a form of coordination without communication:
  each process sets its heap limit
  based on its observations of its own execution,
  without communicating any data with other processes,
  yet the resulting allocation of memory across processes
  nonetheless achieves
  the globally-minimal total garbage collection time.
The lack of communication
  is especially attractive in an untrusted environment,
  like a web browser,
  since no communication means no side channels
  that could compromise the web's strong security properties.
We call the combination of optimality in our framework
  and coordination without communication
  a \textbf{compositional} heap limit rule,
  because multiple such programs can be composed
  without manually tuning the heap limits of each.

We implement our square-root heap limit rule
  in V8, the state-of-the-art garbage-collected JavaScript runtime
  that powers Google Chrome, Microsoft Edge, and other browsers,
  as well as server-side frameworks such as node.js and Deno.
V8's current heap limit computation
  is based around a core multiple-of-live-size rule
  that is modified, overridden, or adjusted
  by multiple heuristics, triggers, and asynchronous processes.
In its place,
  our prototype, MemBalancer,
  monitors allocations in real time
  and adjusts heap limits on the fly
  using a single, uniform implementation
  of the square-root heap limit rule.

We evaluate MemBalancer
  on six real-world web applications:
  Facebook, Gmail, Twitter, CNN, Fox News, and ESPN.
MemBalancer achieves a better trade-off
  between average heap usage
  and garbage collection time
  than V8's current heap limit rule.
Depending on how MemBalancer is tuned,
  an average reduction
  of \BROWSERIIMemorySaving of average heap usage
  or \BROWSERIISpeedup of total garbage collection time
  is achievable.
On memory-intensive JavaScript benchmarks
  from the ACDC~\cite{acdc} benchmark suite,
  MemBalancer demonstrates reductions
  of up to \ACDCMemorySaving of average heap usage
  without increasing total garbage collection time,
  or \ACDCSpeedup of total garbage collection time
  without increasing average heap usage.

\medskip

\noindent
In summary, this paper contributes:
\begin{itemize}
    \item The concept of a compositional heap limit rule,
        based on a framework that allocates a pool of memory
        across several garbage-collected programs
        (\Cref{sec:theory});
    \item A compositional ``square-root'' heap limit rule,
        which achieves coordination without com\-mu\-ni\-ca\-tion
        and is optimal in our framework
        (\Cref{sec:controlling});
    \item A prototype, named MemBalancer,
        of the compositional square-root heap limit rule
        for the V8 JavaScript engine
        (\Cref{sec:impl}).
\end{itemize}
We evaluate our prototype
  on standard JavaScript benchmarks in \Cref{sec:acdc}
  and on real-world websites in \Cref{sec:websites}.

\section{Case Study}
\label{sec:casestudy}

To understand why compositional heap limit rules
  reduce memory usage and garbage collection time,
  let's examine the behavior
  of a compositional heap limit rule
  on three benchmarks from the JetStream 2
  Javascript benchmark suite:%
\footnote{These three benchmarks
  are the most memory-intensive
  of the 64 total JetStream benchmarks.}
  Typescript, which runs the Typescript compiler
  on a fixed Typescript codebase;
  PDF.js, which renders a fixed PDF file
  with the PDF.js rendering engine;
  and Splay, which creates a splay tree
  and does a sequence of insert and delete operations on it.
All three benchmarks run a kernel in a hot loop,
  making it easy to understand their garbage collection behavior.
Notably, the PDF.js benchmark contains a memory leak,
  meaning that it consumes more and more memory over time.%
\footnote{
  To our knowledge this was first pointed out by \citet{vm-hot-cold}.
}
This memory leak causes increasing mis-allocation
  between the three benchmarks
  as the iteration count increases.%
\footnote{
  Sadly, memory leaks are also common in real-world websites.
}

\begin{figure}
    \centering{\fbox{
        ${\color{blue!90!black}\bullet}$ TypeScript \quad
        ${\color{orange!60!brown}\bullet}$ Splay \quad
        ${\color{green!60!black}\bullet}$ PDF.js
    }}
    \includegraphics[width=0.5\linewidth]{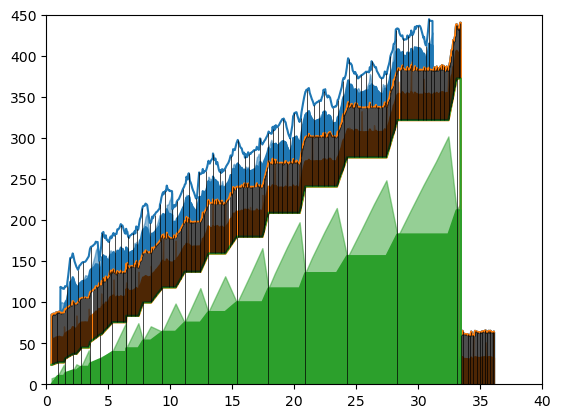}%
    \includegraphics[width=0.5\linewidth]{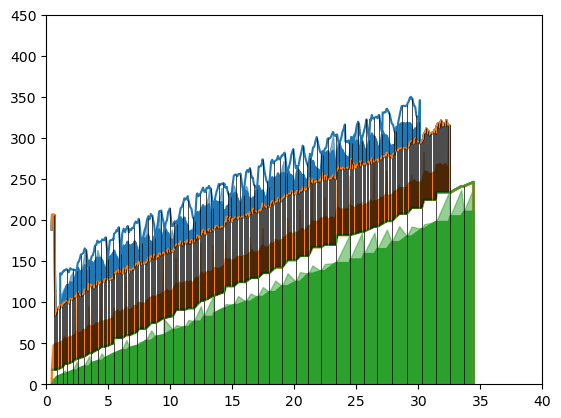}
    \caption{Heap limit and usage plots
      for the three JetStream 2 benchmarks
      using either V8's current heap limit algorithm
      (on the left)
      or MemBalancer (on the right).
    Time runs along the horizontal axis (in seconds)
      and memory use along the vertical axis
      (in megabytes).
    Both plots use the same axes;
      the MemBalancer run (on the right)
      uses less memory and finishes faster.
    Each benchmark is shown in a different color
      (as described by the legend)
      and consists of three values:
      the live memory (dark color),
      current heap memory usage (light color)
      and current heap limit (white).
    A black line is drawn on each thread
      when it collects garbage.
    MemBalancer allocates less memory to PDF.js
      and more memory to Splay.
    Because Splay collects garbage very often,
      this trade is profitable,
      reducing garbage collection time and memory usage.
    }
    \label{fig:js2tl}
\end{figure}

\Cref{fig:js2tl} illustrates the memory allocation behavior
  of two runs of this benchmark,
  one using V8's current heap limit rule,
  and one using MemBalancer,
  our implementation of a compositional heap limit rule
  for V8.
In each plot,
  the different colors represent different benchmarks,
  showing their memory use over time.
The MemBalancer run uses 3.5\% less memory,
  yet spends 20\% less time garbage-collecting.
This is because V8's current heap limit rule
  is not compositional:
  it overallocates memory to some benchmarks
  and underallocates memory to others.
Specifically,
  it allocates minimal memory
  to the Splay and TypeScript benchmarks,
  causing rapid garbage collections,
  while PDF.js is allowed a much larger heap.
At its core, this is because V8's current heap limit rule
  allocates memory proportionally to
  current live memory.
Due to the memory leak,
  PDF.js has a large amount of live memory,
  and is thus allocated a large heap,
  even though it doesn't use this heap particularly effectively.
In the right plot,
  MemBalancer allocates less memory to PDF.js
  and more to Splay,
  dramatically reducing overall garbage collection time
  without much affecting overall memory use,
  due to its use of a compositional heap limit rule.
Allocating slightly more memory to Splay
  and much less to PDF.js
  saves more than enough garbage collection time on Splay
  to compensate for extra garbage collection time on PDF.js.

\begin{table}
    \resizebox{\textwidth}{!}{
    \begin{tabular}{|l|rrr|rrr|rrr|} 
        \hline
        & & & & & Current V8 & & & MemBalancer & \\
        Benchmark & $L$ (MB) & $g$ (MB/s) & $s$ (MB/s) &
        $M - L$ (MB) & GC time (s) & Run time (s) & $M - L$ (MB) &  GC time (s) & Run time (s)  \\
        \hline
        \nameA & \lA & \gA & \sA &  \baseextraA & \basegctimeA  & \baseruntimeA & \mbextraA & \mbgctimeA & \mbruntimeA \\
        \nameB & \lB & \gB & \sB &  \baseextraB & \basegctimeB  &  \baseruntimeB & \mbextraB & \mbgctimeB & \mbruntimeB \\ 
        \nameC & \lC & \gC & \sC & \baseextraC & \basegctimeC &  \baseruntimeC & \mbextraC & \mbgctimeC & \mbruntimeC \\ 
        \hline
        \nameD & \lD & \gD & \sD &  \baseextraD & \basegctimeD  &  \baseruntimeD & \mbextraD & \mbgctimeD & \mbruntimeD \\ 
        \hline
    \end{tabular}}
    \label{tab:js2vals}
    \caption{
      In the first four columns,
        point-in-time estimates of $L$, $g$, and $s$
        for the three JetStream 2 benchmarks,
        taken relatively early
        into a MemBalancer run with $c = \CompareAt\%/MB$.
      In the next six columns,
        measurements of usable heap space $M - L$,
        total garbage collection time,
        and total benchmark run time
        for both MemBalancer and current V8 runs.
      Note that the $M - L$ columns
        for MemBalancer and current V8
        are a point-in-time measurement
        (from similar but slightly different points
         during the benchmark run),
        while the total garbage collection time
        and total benchmark run time
        reflect the whole run.
      MemBalancer allocates the most usable heap space
        to the Splay benchmark,
        while current V8 allocates the most to PDF.js.
      As a result, MemBalancer reduces
        garbage collection time for Splay
        while increasing it for PDF.js,
        resulting in lower total garbage collection time.
    }
\end{table}

To underscore this point,
  \Cref{tab:js2vals} contains
  point-in-time estimates of
  each thread's live memory ($L$),
  allocation rate ($g$),
  and garbage collection speed ($s$)
  for the three benchmarks
  (captured from the MemBalancer run).
The Splay benchmark is unusual
  in both allocating memory and collecting garbage
  much faster than Typescript or PDF.js.
This suggests that Splay should receive
  much more usable heap space
  than PDF.js;
  according to our model (\Cref{sec:theory}),
  roughly $\sqrt{\JSSplayPDFJSl \cdot \JSSplayPDFJSg / \JSSplayPDFJSs} \approx \JSSplayPDFJSExtraMem$ times more.%
\footnote{Note that the numbers in \Cref{tab:js2vals}
  do not exactly match this computation
  due to smoothing.}
However, V8's current heap limit rule
  is based mainly on live memory size
  and allocates the most usable heap space
  to PDF.js benchmark.
This are visible in \Cref{fig:js2tl},
  with PDF.js's portion of the plot becoming shorter,
  with more garbage collections,
  but the TypeScript and Splay portions becoming taller,
  with fewer garbage collections.
Note that Splay still collects garbage
  roughly $\JSSplayPDFJSGCFreq\times$ more often than PDF.js,
  since it allocates memory $\JSSplayPDFJSg\times$ faster
  but only has $\JSSplayPDFJSExtraMem\times$ more extra memory. 
This more-frequent collection is optimal,
  because Splay also allocates much more memory
  than the other two benchmarks.

When we talk about ``allocation'',
  keep in mind that neither current V8
  nor MemBalancer
  have a central controller that allocates space;
  instead, this allocation is the emergent result
  of individual threads making heap limit decisions
  using local heap limit rules.
Of course, both the current V8 heap limit rule,
  and MemBalancer, are configurable;
  V8's current rule has dozens of tweakable parameters,
  which we left at their default values,
  and MemBalancer also has a tweakable $c$ parameter,
  which we chose to be $c = \CompareAt\%/MB$.
Naturally, tweaking these parameters
  could cause V8 to use less memory,
  or alternatively to use more memory
  and thereby spend less time collecting garbage.
However, no matter how these parameters are tweaked,
  because V8's current heap limit rule is not compositional,
  it will always overallocate memory to PDF.js
  and under-allocate memory to the other two threads,
  meaning that there is some parameter
  that causes MemBalancer to use
  both less memory and less time.
That's because the emergent memory allocation
  depends on structural features of the heap limit rule,
  not specific tuning parameters.

\section{V8 Garbage Collector Background}

This section describes the overall architecture
  of the V8 garbage collector
  with a focus on heap limit rules.
V8 is a state-of-the-art JavaScript runtime
  widely used in desktop, mobile, and server-side computing
  to power web applications,
  mobile apps,
  and back-end services.
While Google Chrome is V8's most prominent client,
  desktop applications built with the Electron framework
  (such as Slack, Discord, Spotify, and VS Code),
  mobile applications built with the Android WebView component
  (such as WeChat, Facebook, and Amazon),
  and server-side applications build with node.js and Deno
  all use V8.

Web browser users
  typically use tabs to view multiple websites,
  meaning that a single Chrome instance hosts multiple
  V8 JavaScript runtimes, known as ``Isolates''.%
\footnote{Moreover,
  a single browser tab can host content from multiple sites
  via features like \texttt{<iframe>},
  and a single site can start
  multiple independent and concurrent JavaScript heaps
  using features like WebWorkers.}
These runtimes typically run in different processes%
\footnote{
  In Google Chrome,
    tabs from different \textit{origins}
    (the part of the URL before the path)
    always run in different processes,
    while tabs from the same origin
    may or may not run in different processes.
  Other browsers such as Firefox and Safari
    use fewer processes, but still attempt
    to minimize communication between tabs
    for security reasons.
}
  for reliability, performance isolation, and security.
The security consideration
  has recently become more important,
  with CPU vulnerabilities such as Spectre~\cite{SpookJS}
  shown to be exploitable on the web.
Because web pages are untrusted and mutually antagonistic,
  it's  essential that, insofar as possible,
  no communication occurs between Isolates.
Each Isolate therefore contains its own heap
  and makes all decisions,
  including heap limit decisions,
  independently.

Because web applications are famously memory-hungry,
  web browsers use care\-ful\-ly en\-gi\-neer\-ed,
  high-performance garbage collectors
  tuned for high throughput and low pause times.
V8 uses a generational garbage collector
  with one young generation (nursery)
  approximately 10\,MB in size that uses mark-copy collection,
  and one old generation
  that uses mark-compact collection.
Minor garbage collections
  are triggered when the nursery runs out of space;
  they happen every few seconds.
This paper instead focuses on the old generation
  and major garbage collections.

Major garbage collections in V8 are complex,
  with optimized mark and sweep phases.
To reduce pause times, in addition to a classical stop the world collector,
  V8 uses an incremental marking algorithm
  with a Dijkstra-style write barrier~\cite{Concurrent,ConcurrentBlog}.
This means that the marking phase
  can be run in small chunks
  and interlaced with program execution.
When the program makes modifications to the object graph,
  it updates the GC state atomically
  to ensure that the live/dead state for each object
  (the object's ``color'') is accurate.
Incremental marking phases then propagate those colors
  through the object graph.
Importantly,
  incremental marking means
  that most of the heap
  does not need to be traversed during the
  non-concurrent final marking phase,
  since it already has the correct color.

Incremental marking is run
  concurrently with the mutator thread,
  that is, the JavaScript program itself.
This reduces pause times on the mutator thread
  and is the preferred way to do marking.
However, incremental marking results
  need to be finalized before the compaction phase can start.
This non-incremental, non-concurrent mark phase
  is started when the heap limit is reached,
  but is typically fast because
  most colors are accurate as a result of incremental marking.
This combination
  of concurrent incremental marking
  and non-concurrent non-incremental finalization
  minimizes pause times.

Both the marking and compaction phases
  are also parallelized across multiple threads.
Together with incremental and concurrent marking,
  these garbage collector features
  make it difficult to talk about total garbage collection time.
This paper uses internal V8 APIs to compute the total collection time
  across all incremental marking phases,
  the final non-incremental marking phase,
  and the compaction phase.
These APIs track total CPU time across all threads.

V8 sets old generation heap limits dynamically
  after every garbage collection.
At a high level, V8 has two heap limit rules:
  setting the heap limit to a fixed multiple of live size,
  or adjusting the heap limit over time
  with a target of garbage collection being 3\% of total running time.
The fixed multiple rule forms a cap on the heap limit derived via the 3\% rule.
In the framework of \Cref{sec:theory},
  both of these rules result in a heap limit
  that is directly proportional to live size,
  so neither rule is compositional in our framework.
Many additional factors make minor adjustments
  on top of these two high-level rules;
  for example, laptop and phone builds of V8
  set heap limits slightly differently.
These competing rules and adjustments reflect the fact that
  V8's current memory limit formula
  is not based on any formal model of optimal memory limit,
  and therefore must balance competing concerns
  in a fundamentally ad-hoc way.
  
One challenge with adjusting heap limits after garbage collection
  is that a thread can become idle---%
  for example, a web application can be waiting for user input---%
  while holding on to a large amount of dead memory~\cite{Idle}.
V8 therefore contains an intermittent ``memory reducer'' process
  that attempts to trigger garbage collection during idle times.
Specifically, the memory reducer triggers
  additional major garbage collections
  based on factors like
  a low allocation rate,
  no frame being rendered by the browser,
  and memory pressure being present.
Additionally, the memory reducer
  runs more often in background tabs
  and when a larger portion of the current heap limit is being used.
The memory reducer determines when and how often to fire
  based on a complex boolean formula
  and a state machine that adjusts its behavior over time.

Finally, the Blink process monitors total system memory usage
  to avoid overwhelming system resources.
When system resources are close to being exhausted
  (according to platform-specific
  operating system APIs)
  it notifies each V8 heap via a ``memory pressure'' notification.
Upon receiving such a notification,
  V8 immediately performs a mark-compact garbage collection
  and returns all free pages to the operating system.
In our experiments,
  to avoid noise from memory pressure notifications
  and isolate tabs from each other,
  we use a machine with enough memory
  that memory pressure is never a concern.

Within Google Chrome, V8 is one of several key components.
Another is the rendering engine, Blink,
  which stores web page data
  and draws the web page to the screen.
Blink objects live in their own memory space,
  but can both be referenced from JavaScript
  (such as when a JavaScript object
  stores a reference to an HTML element in the web page)
  and also reference JavaScript objects in turn
  (such as when an HTML element stores a JavaScript callback).
Therefore V8 also has a second garbage collector
  for rendering engine objects,
  which coordinates with the main V8 garbage collector~\cite{Cross}.
While work is ongoing
  to merge the two garbage collectors in V8,
  at the moment they operate largely independently.
This paper focuses on garbage collection
  on the JavaScript heap,
  and so ignores browser-side objects
  except to account for their presence
  when computing heap limits and heap usage statistics.

One thing worth noting is that V8 is optimized
  for low-memory devices such as mobile phones,
  which causes different trade-offs than server-side run-times
  such as JVMs.
For example, in server-side run-times
  the live memory is typically a fraction
  of total heap size to minimize garbage collection time;
  V8 uses much more conservative heap limits,
  and typically keeps the live memory
  to 60--80\% of total heap size.
This means that heap limit rules
  have a numerically-smaller effect on total heap size,
  since most of the heap will typically be taken up
  by live objects.
However, this also suggests that memory is at a premium,
  and therefore that even small reductions in memory use
  are valuable to users.
\section{Compositional Heap Limit Rules}
\label{sec:theory}

Our key insight is that
  every heap limit rule induces an allocation of memory
  across multiple heaps.
With a single heap, 
  all heap limit rules are pareto-optimal
  because they all merely pick a point
  that trades off memory usage for garbage collection time;
  but with multiple heaps, the induced allocation need not be pareto-optimal.
A \textit{compositional} heap limit rule 
  induces a pareto-optimal allocation of memory
  across multiple programs.

To derive a compositional heap limit rule theoretically,
  consider a simple model
  of multiple garbage-collected language runtimes
  running simultaneously on a single computer.
For concreteness, imagine that it is
  a multi-tab browser where each tab runs JavaScript
  and has a stop-the-world mark-compact garbage collector.
In our model,
  each browser tab runs JavaScript that allocates memory
  at a fixed rate until that tab's heap limit is breached.
At that point, JavaScript execution stops
  and garbage collection runs.
While V8's mark phase can run concurrently with JavaScript,
  this model is still a close match
  because, thanks to incremental marking,
  mark phases are typically fast
  and the stop-the-world compaction phase dominates.
  
\begin{wrapfigure}{r}{2.5in}
\begin{tabular}{ll}
 Variable & Meaning \\\hline
 $L$ & Live memory \\
 $g$ & Allocation rate \\
 $s$ & Garbage collection speed \\
 $M$ & Heap limit
\end{tabular}
\caption{Variables used in our theoretical model.}
\end{wrapfigure}

Consider a single garbage collection cycle
  that starts with $L$ bytes of live memory 
  and a heap limit of $M$ bytes.
Throughout the cycle,
  JavaScript first runs for $t_m$ seconds,
  followed by $t_g$ seconds of garbage collection.
If JavaScript allocates memory at an average rate of $g$,
  then in $t_m$ seconds it allocates $g t_m$ memory,
  which must be equal to $M - L$.
Assume furthermore that
  the garbage collector's running time is proportional to $L$
  at a fixed garbage collection speed $s$;
  this is, again, a rough match to V8's compaction phase.

Putting these assumptions together,
  this garbage collection cycle
  involves $t_m = (M - L) / g$ seconds of JavaScript run time
  and $t_g = L / s$ seconds of garbage collector run time.
To total this up across multiple heaps,
  which start and stop each garbage collection cycle
  at different times,
  we amortize garbage collection time
  over the full length of the cycle.
Of our overall running time,
  a $t_g / (t_m + t_g)$ fraction
  is spent in garbage collection.
If we assume that $t_m \gg t_g$,
  this simplifies to:%
\footnote{
  This assumption can be dropped, and
    the resulting heap limit rule is very similar;
    it is merely the square root rule derived in this paper,
    multiplied by a factor of $s / (s + g)$.
  Since garbage collection speed is faster than the allocation rate,
    this factor is typically close to 1,
    and dropping it does not impact the heap limit much.
}
\[
  \mathit{ratio} := \frac{t_g}{t_m} = \frac{L}{s} \frac{g}{M - L}
\]
Note that $s$, $g$, and $L$ are all
  program-specific parameters that cannot be controlled directly.
However, $M$, the heap limit, is an arbitrary parameter.

The goal is now to choose $M$ optimally.
Since the browser has multiple tabs,
  the objective
  is to minimize the sum $\sum\mathit{ratio}$ across all the tabs
  by optimally choosing each tab's heap limit $M$.
To do so, consider the derivative of $\mathit{ratio}$
  with respect to $M$.
This derivative $\partial \mathit{ratio} / \partial M$
  represents the ``exchange rate''
  between a larger heap limit
  and a lower percentage of running time spent on garbage collecting;
  it is measured in $\%/MB$ or some equivalent unit.
For $\sum\mathit{ratio}$ to be minimized,
  a kind of ``no-arbitrage condition'' needs to hold:
  reducing one tab's heap limit by a tiny amount
  and then increasing another tab's heap limit by the same amount
  should not impact $\sum\mathit{ratio}$.
Therefore, $\partial \mathit{ratio} / \partial M$
  must be equal for all tabs.%
\footnote{
This optimization problem can be solved more formally
  using the method of Lagrange multipliers.
}
Taking this derivative symbolically yields
\begin{equation}
\label{eq:constraint}
-\frac{\partial \mathit{ratio}}{\partial M} = \frac{L g}{s (M - L)^2}
\end{equation}
For a heap limit rule to be compositional,
  this must equal the same constant $c$ for all heaps.

\textbf{Standard heap limit rules are not compositional.}
Specifically, if $M$ is set
  to a multiple $(\alpha + 1) L$ of live memory,
  then $c$ is equal to $g / s \alpha^2 L$,
  which differs for heaps
  with different $g$, $s$, and $L$.
Of course, for a single heap, tuning $\alpha$ appropriately
  can achieve any heap limit.
However, multiple heaps with the same $\alpha$
  can only achieve certain combinations of heap limits,
  and those combinations do not include
  the combination that minimizes total garbage collection time.
This demonstrates
  how considering multiple heaps
  places strong constraints on heap limit rules.

Returning to \Cref{eq:constraint},
  solving for $M$ in terms of $c$ yields:
\begin{equation}
\label{eq:sqrt-rule}
  M = L + \sqrt{L g / c s}
\end{equation}
Amazingly,
  all parameters involved in \Cref{eq:sqrt-rule}---%
  the garbage collection speed $s$,
  the allocation rate $g$,
  and the live memory $L$---%
  are local to a single heap.
\Cref{eq:sqrt-rule} therefore defines
  a heap limit rule---%
  one that ``coordinates without communicating'',
  achieving an optimal allocation of memory across heaps
  without any communication between those heaps.
This ``square-root'' rule has the expected heap limit behaviors:
\begin{enumerate}
\item Tabs that have more live memory (with large $L$)
  should have more ``extra memory'' $M - L$,
  since each garbage collection has a higher ``fixed cost''
  from traversing the live memory.
\item Tabs that allocate more quickly (with large $g$)
  should have a higher memory limit,
  since they hit that limit more often.
\item Tabs whose heaps can be mark-compacted more quickly (with large $s$)
  should have a lower memory limit,
  since garbage collections are less costly.
\end{enumerate}
In other words,
  as expected,
  an optimal memory limit
  allows more memory to tabs that
  need more memory, produce garbage faster,
  and collect garbage slower.

But crucially, the square-root heap limit rule
  allocates ``extra memory'' $M - L$ to each tab
  proportional not to $L$, $g$, and $s$,
  but to their square-roots.
This is the key to minimizing
  total garbage collection time across multiple heaps.
Compared to standard proportional heap limit rules,
  the square-root rule
  results in lower heap limits for heaps with large live memory (high $L$), 
  and higher heap limits for heaps with small live memory (low $L$);
  in other words, it is \textbf{sublinear}.%
\footnote{
  Firefox's TraceMonkey JavaScript runtime
    uses a proportional heap limit rule,
    but with smaller constants of proportionality
    as the live memory increases.
  This potentially approximates
    the optimal square-root rule behavior.
}
Heaps that allocate a lot (high $g$)
  or are slow to collect garbage (low $s$)
  also get higher heap limits with the square-root rule,
  since it takes these factors into account.
Of course, the $c$ parameter can be changed
  to increase or decrease overall memory usage,
  just like tuning the constant of proportionality
  in a proportional heap limit rule.

In practice,
  a compositional heap limit rule
  can also improve performance
  for a single heap.
In our model, program behavior ($g$, $L$ and $s$)
  is invariant over time.
However, real-world programs change their behavior over time,
  such as first loading data and then executing.
A standard, proportional heap limit rule
  will give too high a heap limit when $L$ is large
  and too small a heap limit when $L$ is small.
A compositional heap limit rule,
  by contrast,
  will minimize total garbage collection time
  across the full program execution.

This paper focuses on the square-root rule
  where $c$ is a fixed constant.
That said, \Cref{eq:constraint} allows $c$ to vary
  as long as it is the same for all heaps at any given point in time.
One could imagine a centralized controller
  that adjusts $c$ up and down
  to use all available system resources
  without causing swapping, as in \citet{Advise}.
Or, $c$ could vary
  based on whether or not a phone is charging or using battery.

Another possibility is to optimize for something
  other than total garbage collection time.
For example, garbage collection time on different heaps
  can be \textit{weighted},
  with background tabs having a lower weight.
  Then the $c$ value for a tab
  should be proportional to one over each tab's weight.%
\footnote{
  The derivation is similar to that
    for \Cref{eq:sqrt-rule}.
}
These options show that, despite its strictness,
  \Cref{eq:constraint} admits a lot of flexibility.
\section{A Square-Root Rule Algorithm}
\label{sec:controlling}

At a high level,
  implementing the square-root heap limit rule
  of \Cref{eq:sqrt-rule}
  is straightforward:
  measure $L$, $g$, and $s$ for each tab
  and compute $M = L + \sqrt{Lg/cs}$ from those values.
However, two challenges need to be resolved for
  to put this idea into practice.

The first challenge is defining and measuring
  the live memory $L$,
  average allocation rate $g$,
  and average garbage collection speed $s$.
The challenge is that, while $g$ is observable in real time
  (thanks to a simple counter incremented by the allocator),
  $s$ and $L$ are not.
Handling this requires a multi-threaded design.
Because $g$ can be observed in real time,
  a ``heartbeat'' thread wakes once a second
  to record the current allocation rate $g$
  and re-compute the heap limit $M$.
This can raise or lower the heap limit;
  if the heap limit is lowered below
  the current heap size,
  a garbage collection will be triggered later by V8.
By contrast, $L$ and $s$
  are not observable in real time,
  and can only be measured accurately
  immediately after a garbage collection.%
\footnote{
Actually, due to incremental GC,
  multiple inter-related heaps, and other details of V8,
  a garbage collection may not clear out all dead objects,
  meaning that $L$ might not be precisely observed.
In practice, however, the memory left after garbage collection
  is mostly live objects,
  so makes for a good-enough estimate of $L$.
}
Our proposed approach, which we call MemBalancer,
  therefore measures $L$ and $s$ on the main thread
  after every major garbage collection,
  recording $L$ and $s$ and recomputing the heap limit $M$.
Major garbage collections also measure and update $g$
  to avoid measuring $g$ across major GC boundaries.
As a result,
  heap limits are updated at least once per second,
  or more often if garbage collections are frequent.

The second challenge is prediction.
In the model of \Cref{sec:theory},
  the $s$ value in \Cref{eq:sqrt-rule}
  represents the time to collect garbage
  at the \textit{end} of the current cycle,
  so it is fundamentally a predictive value.
Similarly, $g$ measures the average allocation rate
  over the current cycle,
  so is again a predictive value.
Moreover, both $s$ and $g$ change as the program executes,
  so their estimated values need to react to changes;
  but both are measured based on run times,
  so need to dampen noise.
MemBalancer therefore smooths the new allocation data
  using an exponentially weighted moving average,
  which estimates a value as a weighted linear combination
  of the current value and the last estimate.
For $g$,
  a smoothing constant of $\alpha_g = \GarbageRateDecayPerSec$ is used,
  so that the half-life of any data point
  is about thirteen seconds,
  which reduces jitter enough to prevent spurious garbage collections
  but still allows the rate $g$ to respond rapidly
  to changes in program behavior.
Similarly, the measured $s$ is smoothed,
  with a smoothing constant of $\alpha_s = \GCSpeedSmoothingPerSample$,
  to account for anomalously long garbage collections.
Both $g$ and $s$ are measured
  by dividing a number of bytes by a duration;
  as is standard, we smooth the top and bottom individually.
\section{Implementation in V8}
\label{sec:impl}

We instantiated this high-level algorithm
  for the V8 JavaScript virtual machine,
  basing our prototype on a stable version of V8
  from 7 June 2022 (commit hash \texttt{b1413ed7}).
Broadly speaking,
  our V8 prototype of MemBalancer
  follows the high-level algorithm in \Cref{fig:membalancer_impl}.
The V8 \texttt{Heap} constructor
  creates an additional heartbeat thread for each heap.
This heartbeat thread wakes once per second
  to record the current heap memory usage
  (using the \texttt{SizeOfObjects()} API in V8)
  and time since last update.
This is compared to the previous record
  (stored in new \texttt{Heap} member fields)
  which is smoothed
  and written to atomic member fields on the \texttt{Heap},
  as in \F{on\_heartbeat()} in \Cref{fig:membalancer_impl}.
Every time a major garbage collection occurs,
  the main thread uses the existing \texttt{GCTracer} API
  to measure live memory $L$
  and garbage collection speed $s$.
Garbage collection speed is smoothed
  and then both values are written
  to two other atomic member fields on the \texttt{Heap},
  as in \F{on\_gc()} in \Cref{fig:membalancer_impl}.
The member fields are atomic
  to ensure reliable communication between the two threads;
  using atomic variables instead of a mutex
  is important for keeping pause times low.
After either kind of update,
  the heap limit is recomputed
  following \F{compute\_M()} in \Cref{fig:membalancer_impl}.
Because each Isolate has its own heartbeat thread,
  multiple heaps in a process
  (such as multiple WebWorkers)
  do not communicate any information
  and are performance-isolated.

We use environment variables to enable MemBalancer 
  and set the $c$ parameter.
This is convenient for experiments,
  but in a production implementation,
  these values could be hard-coded,
  set by compilation flags,
  or passed to V8 by Blink
  through V8's options API
  or through Blink's Performance Manager API.
For example, ``V8 Lite'' builds,
  intended for low-memory environments,
  could use larger values of $c$
  to trade more garbage collection time for less memory use.
All told, our prototype of MemBalancer
  requires roughly 200 lines of modifications to V8 
  (plus a couple dozen lines of logging code).

\begin{figure}
    \begin{minipage}{0.4\linewidth}
    \setlength{\jot}{0pt}
    \begin{align*}
      & s_m^*, s_t^*, L^*, g_m^*, g_t^* = 0, 0, 0, 0, 0 \\[6pt]
      & \K{def}\:\F{on\_gc}(s_m, s_t, L): \\
      & \> s_m^* = \alpha_s s_m^* + (1 - \alpha_s) s_m \\
      & \> s_t^* = \alpha_s s_t^* + (1 - \alpha_s) s_t \\
      & \> L^* = L \\[6pt]
      & \K{def}\:\F{on\_heartbeat}(g_m, g_t): \\
      & \> g_m^* = \alpha_g g_m^* + (1 - \alpha_g) g_m \\
      & \> g_t^* = \alpha_g g_t^* + (1 - \alpha_g) g_t \\[6pt]
      & \K{def}\:\F{compute\_M}(M): \\
      & \> E = \sqrt{\frac{L^*}{c}\frac{g_m^* / g_t^*}{s_m^* / s_t^*}} \\
      & \> M = L^* + \operatorname{max}(E, E_{min}) + M_{nursery} \\
    \end{align*}
    \end{minipage}%
    \hfill%
    \begin{minipage}{0.55\linewidth}
    \begin{tabular}{ll}
        Input & Meaning \\\hline
        $L$ & Live memory after collection complete \\
        $s_m$ & Bytes collected during collection \\
        $s_t$ & Collection duration \\
        $g_m$ & Memory allocated since last heartbeat \\
        $g_t$ & Time since last heartbeat \\[6pt]
        Variable & Meaning \\\hline
        $x^*$ & Smoothed estimate of variable $x$ \\
        $c$ & MemBalancer tuning parameter \\
        $M$ & Computed heap limit \\[6pt]
        Constant & Meaning \\\hline
        $\alpha_g$ & $.95$, smoothing factor for $g^*$ \\
        $\alpha_s$ & $.5$, smoothing factor for $s^*$ \\
        $E_{min}$ & $2\MB$, minimum extra heap space \\
        $M_{nursery}$ & $10\MB$, nursery size \\
    \end{tabular}
    \end{minipage}
    \caption{
    On the left, pseudocode for the MemBalancer algorithm.
    On each major garbage collection,
      \F{on\_gc()} is run after the collection,
      followed by \F{on\_heartbeat()} and then \F{compute\_M()}.
    Once a second on a heartbeat thread,
      \F{on\_heartbeat()} is run,
      followed by \F{compute\_M()}.
    On the right, the meaning of
      all of the inputs, variables,
      and constants used by MemBalancer.
    }
    \label{fig:membalancer_impl}
\end{figure}

Internal V8 invariants
  required minor changes to \Cref{eq:sqrt-rule}.
V8 doesn't quite garbage collect
  when the heap limit is reached;
  instead, it garbage collects
  when the remaining space on the heap
  would be less than the nursery size.
We therefore add the nursery size, \TotalMemoryFloorInMB\MB{} to the heap limit computed by MemBalancer
  before setting it as the V8 heap limit.
We also noticed that,
  since web applications were event-driven,
  they typically had long periods of idleness
  in between computations.
During an idle period, no memory is allocated,
  meaning that the estimated $g$
  decreases smoothly toward 0.
When another event occurs,
  the heap limit is breached immediately,
  causing a garbage collection.
This could cause an out-of-memory situation
  if $g$ is low enough that that garbage collection
  does not significantly raise the heap limit.
We therefore added
  a minimum usable heap memory limit $E_{min}$
  to MemBalancer,
  meaning that its heap limit
  is always set to at least the live memory $L$
  plus \MinHeapExtraSizeInMB\MB.
When a new event comes in,
  MemBalancer has enough time
  to measure a new $g$ value and update the heap limit
  before it is breached.
While these tweaks mean that MemBalancer
  is not a strictly compositional heap limit rule,
  the excess is only a few megabytes per tab,
  much smaller than other per-tab fixed costs within the browser
  and similar to V8's current heap limit rule.%
\footnote{V8 currently ensures, for example,
  that all heaps have at least 2MB of heap memory unused,
  increasing that to 8MB as memory becomes available.}

Thanks to MemBalancer's principled heap limit rule,
  our prototype can remove V8 components that patch over issues 
  with a purely proportional heap limit rule.
A common pattern in web applications is that
  the application idles until a user event occurs.
Once user input occurs,
  the application runs, allocating and using memory,
  until it is done responding to the event
  and returns to idleness.
Applications also make network requests
  and idle until a response comes back.
Such applications need to be garbage-collected
  to ensure they don't needlessly hold on to garbage.

In V8 currently,
  the heap limit is only updated
  when a major garbage collection occurs.
To avoid idle heaps retaining memory,
  V8 thus has a separate memory reducer process
  that periodically triggers garbage collections
  when a thread is idle and not allocating~\cite{Idle}.
However, tuning this memory reducer process is difficult;
  if it runs too often, it wastes time collecting garbage
  from only temporarily-idle threads,
  while if it runs less often threads retain garbage for too long.

By contrast, MemBalancer adjusts the heap limit
  at a steady one-second cadence
  as its estimate of $g$ changes,
  much more often than the memory reducer is triggered.
When a thread goes idle,
  its estimated $g$ decays exponentially
  (due to the exponentially weighted moving average)
  and thus so does its extra memory.
This quickly leads to a garbage collection,
  collecting garbage memory from the idle thread.
This supersedes the memory reducer, which we could then remove.
MemBalancer triggers these idle-time garbage collections
  earlier and less frequently than the memory reducer would,
  leading to lower memory usage
  and lower garbage collection time.

To aid evaluation of V8 heap limit rules,
  we built a garbage collection logging system for V8.
Each \texttt{Heap} writes to a private log file
  every time a major garbage collection occurs
  and every time the heartbeat thread measures $g$.
This allows reconstructing allocation and garbage collection behavior.
Because it involves access to the file system,
  this logging system requires turning off the V8 sandbox
  and would not be advisable in a production-ready implementation.
Aggregated log files could instead be made available
  through the browser's built-in developer tools,
  potentially giving web developers
  granular information about $L$, $g$, and $s$
  that they could use to further improve
  the application's performance.
\section{Evaluation}

We evaluate MemBalancer
  on the ACDC-JS garbage collection microbenchmark
  and on real-world websites,
  demonstrating substantial reductions
  in garbage collection time
  and average heap usage.
  
\subsection{Microbenchmarking with ACDC-JS}
\label{sec:acdc}

We evaluate MemBalancer
  against V8's current heap limit rule
  using the ACDC-JS benchmark suite~\cite{acdc}.
ACDC-JS is an existing JavaScript garbage collection benchmark
  that simulates real-world heap shapes.
The heap shape is chosen via a set of tuneable parameters,
  including object liveness and object size,
  which affect our $L$ and $s$ parameters.
(ACDC-JS's other parameters affect
  structural properties of the object graph
  with less direct impact on MemBalancer.)
Given these parameters,
  ACDC-JS allocates and deallocates objects in a tight loop,
  so that almost all run time is spent in
  the JavaScript runtime's allocator and garbage collector.
It thus makes a challenging test of MemBalancer's impact
  on program run time.
For our test
  we chose four threads corresponding to
  (object size, object liveness) of (8, 1) (8, 16) (64, 8), and (64, 128).
We tune each run to take approximately 1 minute
  and leave all other parameter values at their default.

We evaluate both MemBalancer and current V8
  on total garbage collection time
  and average heap memory usage.%
\footnote{
We run our experiments on a machine
  with an i7-5820k 12-core CPU (at 3.30GHz)
  and 16GB of DDR4 memory
  running Zorin OS 16.1,
  a Linux distribution based on Ubuntu 20.04.2 LTS.
}
Garbage collection time is measured
  using the standard V8 \texttt{GCTracer} API.
Only major garbage collection time is included,
  and for incremental garbage collection,
  both the concurrent, incremental mark phase
  and the finalization phase
  are measured and added together.
Heap memory usage uses
  the V8 \texttt{SizeOfObjects()} and \texttt{External\-Memory\-Allocated\-Since\-Mark\-Compact()}
  APIs to track total object size in the old generation.
This definition of heap memory usage
  does not account for fragmentation,
  though we expect the effect from this to be small
  except in cases of extremely low memory,
  because V8 uses a mark-compact garbage collector
  that keeps fragmentation low.
Non-heap-allocated objects like jitted code or feedback vectors
  are not included in this measure.
Heap memory usage is measured once per second on each V8 isolate
  and all measurements are reported and averaged
  to compute the average heap memory usage.
Both garbage collection time and heap memory usage
  are then summed across the four threads.

Comparing MemBalancer to current V8 requires care
  because different $c$ values will cause MemBalancer
  to have different average heap usage
  and garbage collection time.
We therefore compare the current V8 heap limit algorithm
  (the baseline)%
\footnote{
Note that current V8 does not have
  an easily tunable parameter for garbage collection frequency,
  so each run of the baseline produces a single measurement,
  not a trade-off.
}
  against MemBalancer with different values of $c$
  ranging from $\ACDCMinC$ to $\ACDCMaxC$.
MemBalancer is superior to current V8
  if the range of heap usage / collection time trade-offs
  achievable by MemBalancer
  is superior to the heap usage and collection time
  of current V8.

\begin{figure}
\begin{minipage}{0.5\linewidth}
\includegraphics[width=\linewidth]{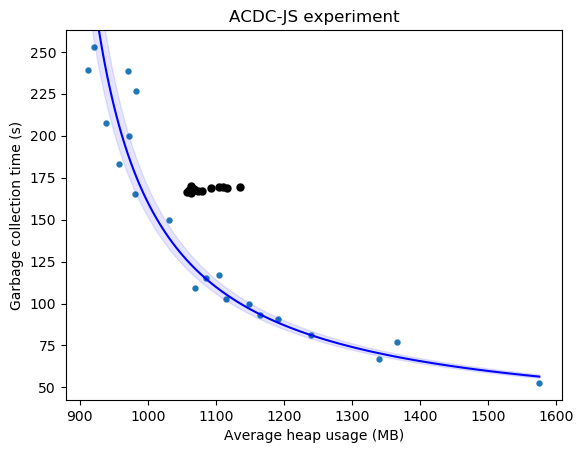}
\end{minipage}%
\hfill%
\begin{minipage}{0.48\linewidth}
\caption{
  MemBalancer and current V8 on ACDC;
  Each point on the scatter plot corresponds
    to a single run of the ACDC-JS experiment;
    blue points use MemBalancer
    (with different $c$ values)
    and black points use current V8
    (with identical parameters).
  Some MemBalancer runs strictly dominate all current V8 runs,
    meaning that MemBalancer achieves
    a better trade-off than current V8.
  The relatively tight dispersion of black dots
    shows that measurement noise has little effect on the results.
  \vspace{.6in}
}
\label{fig:ACDC}
\end{minipage}
\end{figure}

The results are shown in \Cref{fig:ACDC},
  which shows garbage collection time and average heap usage
  for MemBalancer (blue)
  and current V8 (black).
To estimate the improvement in average heap usage,
  we also estimate a regression line
  based on the model in \Cref{sec:theory}.
\Cref{fig:ACDC} shows the regression line
  as well as a 95\% confidence interval
  (two standard errors)
  around the regression line.
  
Depending on the value of the parameter $c$,
  MemBalancer either has
  less average heap memory usage,
  less total garbage collection time,
  or both.
This clearly indicates that on these benchmarks,
  MemBalancer is strictly superior to the baseline.
If a value of $c$ is chosen
  so that MemBalancer uses
  roughly the same amount of memory as the baseline,
  MemBalancer achieves a reduction
  of about \ACDCSpeedup garbage collection time;
  if $c$ is chosen so that MemBalancer spends
  roughly the same amount of time in garbage collection,
  MemBalancer achieves a reduction
  of about \ACDCMemorySaving average heap usage.

\subsection{Macrobenchmarking with Real-world Websites}
\label{sec:websites}

We demonstrate MemBalancer's effectiveness
  on real-world websites
  by integrating our modified V8 version
  with a Chromium (the open-source project behind Google Chrome)
  checkout from 9 June 2022
  (commit hash \texttt{b13d3f}).
This allowed us to evaluate MemBalancer
  on six popular American websites:
  Facebook, Gmail, Twitter,
  Fox News, CNN, and ESPN.
For each website,
  we wrote simple mock user scripts
  using Pyppeteer, a Python port
  of the browser automation library Puppeteer.
Each user script sends user input to the website
  and then waits for a fixed amount of time
  to allow the website to finish responding to it.
For CNN, Twitter, ESPN, Fox News and Facebook
  our script scrolls down
  by \SingleEvalSCROLLPIX pixels
  every second.
These websites have an ``infinite scroll'' feature,
  meaning that scrolling down causes the website
  to load more news stories from the server
  and render them on the page using JavaScript.
For Gmail, our script opens and closes emails,
  spending \SingleEvalGMAILINBOXTIME seconds in the inbox
  and \SingleEvalGMAILEMAILTIME seconds on each email.
This causes Gmail to repeatedly load and render
  email contents and the inbox,
  both of which require significant JavaScript execution.
It also waits \SingleEvalGMAILWAITTIME seconds upon loading
  to emulate an initial user pause.
For Facebook, we open the Facebook ``Group'' page
  and scroll the infinite scrolling feed on that page.
(We initially planned to test more web pages,
  including Reddit and several more news sites.
 However, these sites banned our computer's IP,
   likely due to anti-bot countermeasures,
   so we had to remove these sites from our evaluation.)
Because these benchmarks take longer to run
  than ACDC-JS,
  we test a narrower range of $c$ values
  between $\WEBMinC$ and $\WEBMaxC$.

Facebook, Gmail, and Twitter
  require log-ins to function.
We manually log in to fresh accounts
  before starting the experiments,
  to ensure that we get past any CAPTCHAs
  and so log-in actions are not measured.
On Twitter and Facebook,
  we chose common topics to ``follow''
  to ensure that the feed contains a lot of content.
We used our Gmail account to register for Twitter and Facebook,
  so it contains many complex emails.
Each website continues taking actions
  until three minutes have elapsed.
Since garbage collection time is typically
  a small fraction of overall website running time
  and our scripts include relatively long waits,
  this results in similar work being done
  no matter the heap limit rule.

Because users typically have multiple tabs open,
  including multiple active tabs,
  our evaluation loads and runs multiple websites simultaneously.
Each run is therefore identified
  by the set of websites opened across tabs.
We test all websites individually,
  all thirty pairs of websites,
  and thirty randomly chosen triples of websites.
When multiple websites are run independently,
  the last website opened is the ``active'' tab.
This distinction is important,
  because inactive tabs do not need to be rendered to the screen
  (an activity that blocks JavaScript from running)
  and also disable or reduce the capabilities of certain APIs
  (for example, timers fire less often).
In the current V8, the memory reducer
  also triggers more garbage collection in inactive tabs.
All tabs use the same $c$ parameter,
  and we run multiple runs of each website group
  with different $c$ values.
For each run,
  we measure average heap usage
  and total garbage collection time.
  
For each benchmark, we use live websites
  and a full Chrome instance,
  meaning that the results are roughly representative
  of real-world usage.
But this also means that some noise is inevitable,
  whether due to live A/B testing
  or different results from ad auctions
  or any other reason.
Chrome is also a complex,
  massively concurrent software system
  causing additional noise.
Unfortunately, eliminating this noise is not possible
  without significantly modifying Chrome,
  which would call into question
  our experiment's external validity.
Therefore we restrict our main research question
  to how often MemBalancer points
  strictly dominate current V8,
  and how often the reverse occurs.
This determination should be less affected by noise.

\begin{figure}
\begin{minipage}{.6\linewidth}
\includegraphics[width=\linewidth]{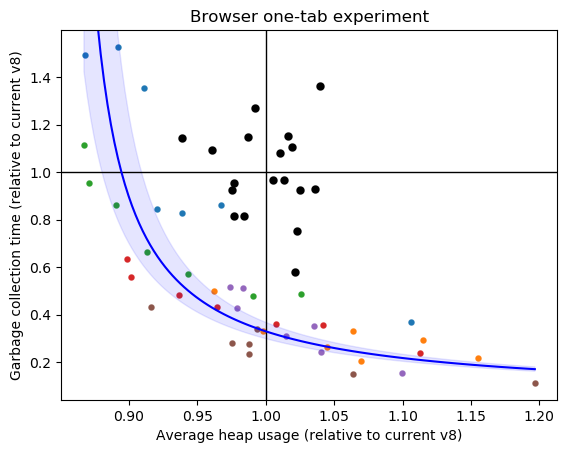}
\includegraphics[width=\linewidth]{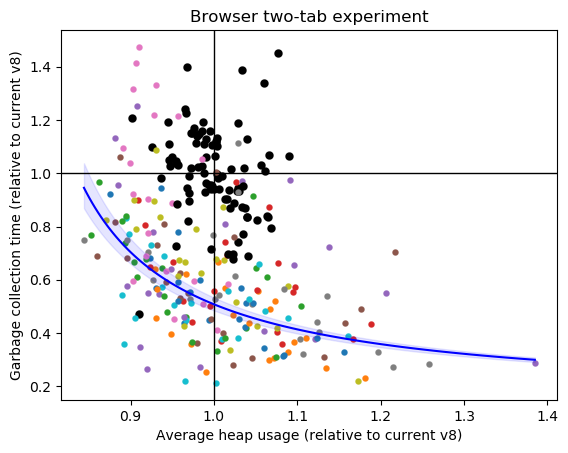}
\includegraphics[width=\linewidth]{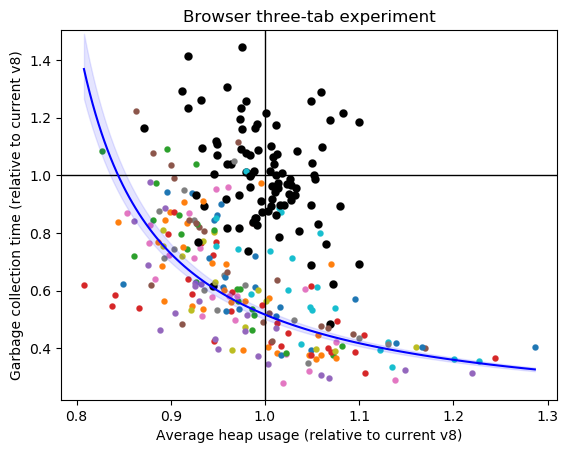}
\end{minipage}%
\hfill%
\begin{minipage}{0.38\linewidth}
\vspace{2.3in}
\caption{
  Results of MemBalancer on six real-world websites,
    with either one, two, or three websites open at a time
    and either MemBalancer (at different $c$ values)
    or current V8 (run multiple times with identical parameters)
    as the heap limit rule.
  Each plot measures
    average heap usage on the horizontal axis
    and total garbage collection time on the vertical axis.
  Both axes are normalized so that the average baseline run
    for a given set of web pages is located at $(1, 1)$.
  Each colorful dot represents a single MemBalancer run,
    with a fixed $c$ value and a fixed set of open web pages;
    each black dot represents a single baseline run
    using current V8 with the default parameters.
  Compared to \Cref{fig:ACDC},
    the black dots are much more widely dispersed,
    reflecting the fact that each run will see
    different network timing, concurrency,
    or page content.
  A regression line
    plus a two-standard-error confidence interval
    is also drawn.
  Almost all of the black dots
    are above and to the right of the regression line,
    meaning that MemBalancer reduces garbage collection time
    and heap memory usage compared to current V8.
  Furthermore,
    MemBalancer's advantage seems to increase
    and the variance seems to decrease
    as the number of simultaneous tabs increases.
}
\label{fig:two-tabs}
\end{minipage}
\end{figure}

Our results are shown in \Cref{fig:two-tabs},
  grouped experiments into three sets of plots:
  for one-tab, two-tab, and three-tab experiments.
As for the ACDC-JS experiments,
  the scatter-plot shows
  average heap usage on the horizontal axis
  and garbage collection time
  on the vertical axis,
  with different colors representing different website combinations.
Additionally, to make different website combinations comparable,
  the current V8 runs are averaged
  and the data is then normalized
  so that that average current V8 run
  sits at $(1, 1)$.

Across all three experiments,
  far fewer points are located in the top-right quadrant
  (where MemBalancer uses more average heap space
   and more garbage collection time)
  than in the bottom-left quadrant
  (where MemBalancer uses less average heap space
   and less garbage collection time).
This demonstrates that MemBalancer is superior
  to V8's current heap limit rule.
We can estimate the relative improvement
  of MemBalancer over current V8
  by estimating a regression line
  based on the data points.
To do so,
  we aggregate every data point from every run
  and use ordinary least squares
  to estimate the trade-off curve,
  under the assumption that it follows the model
  in \Cref{sec:theory}.
The regression line passes
  substantially below the reference point,
  again indicating that MemBalancer
  is superior to V8's current heap limit rule.
Considering the regression line's intersections with the axes,
  we find that MemBalancer saves
  \BROWSERISpeedup of garbage collection time
  or \BROWSERIMemorySaving of average heap usage
  for one tab,
  \BROWSERIISpeedup of garbage collection time
  or \BROWSERIIMemorySaving of average heap usage
  for two tabs,
  and \BROWSERIIISpeedup of garbage collection time
  or \BROWSERIIIMemorySaving of average heap usage
  for three tabs.

Note that the two-tab and three-tab experiments
  achieve larger reductions in average heap usage
  than the single-tab experiments.
With a single tab,
  memory allocation across tabs is a non-issue,
  explaining the lower savings.
But we speculate that the one-tab experiment
  still shows some reduction in average heap usage
  due to inter-temporal trade-offs.
That is, V8's current heap limit computation
  causes a single tab
  to use too much memory (compared to MemBalancer)
  at some points in time,
  and then too little memory (compared to MemBalancer)
  at other points in time.
MemBalancer in effect "trades off across time",
  decreasing both average heap usage
  and total garbage collection time.

\subsection{Threats to validity}
\label{sec:validity}

Both the ACDC-JS and real-world experiments
  demonstrate that MemBalancer
  improves on V8's current heap limit rule,
  achieving a better trade-off
  between average heap usage and total garbage collection time.
Moreover, both sets of experiments
  show a large and significant effect,
  corresponding to a reduction
  of approximately 10\% of average heap usage.
The main threats to the validity of this result
  are noise and representativeness. 

Regarding noise,
  our real-world experiments used live websites
  and thus were subject to variation
  from network timings or even different website contents.
This noise makes it difficult to accurately estimate
  the overall improvement in average heap usage
  and total garbage collection time,
  with our estimates derived from regression lines
  that make strong assumptions about the data.
Nonetheless,
  even the real-world experiments
  clearly have more points
  where MemBalancer strictly dominates current V8
  than vice versa, demonstrating that the effect is positive.
Moreover,
  the ACDC-JS experiment avoids most of these sources of noise,
  with both current V8 and MemBalancer points closely spaced,
  demonstrating that noise does not drive the results.

Regarding representativeness,
  the ACDC-JS experiment is based on a microbenchmark
  that spends its time almost exclusively
  in allocation and garbage collection.
However, our real-world website evaluations
  use websites responsible for billions of daily active users,
  and our user scripts execute common user actions.
Naturally, the set of evaluation websites could be expanded,
  and the universe of web pages is extremely diverse.%
\footnote{
  Though note that, due to anti-bot scripts and other protections,
  evaluating on real-world websites is challenging.
}
That said, the combination of real-world and microbenchmark results
  already strongly supports the conclusion
  that MemBalancer is superior to V8's current heap limit rule
  in practical applications.

Finally,
  our results evaluate MemBalancer and current V8
  on average heap usage and total garbage collection time.
However, users may have more complex desires.
For example, users may care less
  about garbage collection time on background tabs.
MemBalancer could accommodate this preference
  by using a larger $c$ value for background tabs,
  but we do not evaluate this.
Users may also have additional preferences
  around peak memory usage,
  maximum garbage collection pause times,
  fragmentation,
  or other qualities not controlled by MemBalancer.
Modeling and integrating these into MemBalancer
  is a possible direction for future work.
\subsection{Deploying MemBalancer}
\label{sec:discussion}

\paragraph{Firefox}

MemBalancer is likely applicable to JavaScript runtimes
  besides V8.
For example, the Firefox SpiderMonkey JavaScript engine,
  like V8, uses a collection of heuristics
  to choose its heap size.
In Firefox, a thread can use one of two different
  heap limit rules:
  a ``high frequency'' limit
  (when the garbage collector is run often)
  and a ``low frequency'' limit.
Moreover,
  while the heap limit rule
  works by multiplying the live memory by a fixed constant
  (proportional),
  the constant of proportionality decreases
  as the live memory increases.
In other words, the heuristics used by Firefox
  already approximate some structural features
  of MemBalancer.
Moreover,
  the mix of heuristics introduces problems;
  for example, allocating slightly more or less often
  can cause a program to switch from
  the low-frequency to the high-frequency rule
  due to the hard cut-off between them.
The developers believe this is a cause
  of undesired garbage collection behavior~\cite{our-bug}.
The Firefox developers have implemented MemBalancer
  and are testing its effects on Firefox's performance
  with the aim of making it the default heap limit algorithm.

\paragraph{Racket}
We sent a preprint on this paper
  to Matthew Flatt, a core developer
  for the Racket programming language~\cite{racket}.
He implemented a simplified version of the square-root rule
  that sets $M = L + c \sqrt{L}$
  and tested it on his preferred Racket benchmark,
  a single-threaded batch-mode build
  of Racket and related libraries and documentation.
Previously, Racket used a simple proportional rule
  with $M = 2 L$.
The simplified square-root rule
  lowers memory use by 10\% while incurring a 1-2\% slowdown.
Note that multi-threaded or multi-process Racket programs,
  which weren't directly evaluated,
  would likely show larger gains.
The Racket developers also tested a version of the square root rule
  more similar to MemBalancer,
  including smoothed measurements of $g$ and $s$,
  but the results were no better.
That could be because
  Racket program execution is more uniform than web pages.
Similarly, the Racket version of the square root rule
  did not use a heartbeat threads
  because idleness is less common for Racket programs.
The new square-root rule heap limit
  was merged into mainline Racket~\cite{racket-commit}
  and released as part of Racket 8.5.

\section{Related Work}

Fast and efficient garbage collection
  has been studied for decades.
Garbage collection was invented in 1959
  for the Lisp programming language~\cite{lisp}.
The canonical survey of the field
  is the Garbage Collection Handbook~\cite{gcBook}.

The first garbage collector,
  for the Lisp programming language,
  limited the heap to
  15\thinspace000 \texttt{cons} cells
  and triggered garbage collection
  when this heap was exhausted~\cite{lisp}.
Already, the trade-off between heap memory usage and garbage collection is noted:
  ``[garbage collection's] efficiency
  depends upon not coming close to exhausting
  the available memory with accessible lists''~\cite{lisp}.
At the time, the heap limit simply reflected available core memory
  (time-sharing operating systems not yet being in common use)
  and so was not configurable by the user.
Today, of course, garbage-collected languages
  often allow the user to configure the heap limit
  to maximize performance
  in the presence of other programs
  running on the same machine.
Such manual tuning may be necessary
  to achieve maximal performance~\cite{Uber}
  but is complex and requires substantial expertise.

The system most similar to MemBalancer
  is \citet{Advise}'s ``advice server'',
  published decades ago.
The advice server uses a model of garbage collection
  similar to that in \Cref{sec:theory}
  to estimate the impact of different heap limits,
  and adjusts heap limits across multiple heaps
  to reduce total garbage collection time.
Like MemBalancer, the advice server
  measures $g$ and $s$ during program run time,
  and they note a similar square-root dependence.
Unlike MemBalancer, heap limits are changed
  only after garbage collections,
  and smoothing for $g$ and $s$ is not investigated.
This means the advice server requires longer
  program run times---30 minutes instead of 3 minutes.
More importantly, the advice server is a centralized server
  that attempts to adjust overall memory usage
  to consume all available system resources.
This requires a fully trusted environment.
The notion of a compositional heap limit rule, however,
  sidesteps the need for a centralized server,
  replacing it with coordination without communication.
If matching overall memory usage to system resources is desired,
  the advice server could advise processes
  on the choice of $c$ instead of
  directly controlling memory usage.
We do not know of any current uses
  of the ``advice server'' model.

Modern garbage-collected languages
  have automatic heap limit rules
  to adjust the heap limit based on program behavior.
Typically, the heap limit is set to be
  a multiple of live memory
  (or, equivalently, the fraction of time
  spent on garbage collection
  is held to a constant)~\cite{Oracle}.
\citet{PerfGC} analyze such rules
  for a collection of different garbage collection algorithms
  and find that multiples as large as
  $3\times$ or $5\times$ are necessary
  to reduce the overhead of garbage collection
  to negligible levels, though modern garbage collectors typically use lower multiples.

Since greater heap limits improve performance,
  determining the maximum possible heap limit is useful.
Several papers introduce dynamic heap limit rules
  to determine the greatest heap limit
  that does not cause paging~\cite{Resizing,Program,Control,Richard}.
Avoiding paging is particularly important
  because garbage collection
  can trigger worst-case behavior
  in paging algorithms,
  though garbage collectors can be designed
  to avoid this problem~\cite{NoPaging}.
Another line of work introduces a central controller
  that can make allocation decisions across multiple programs.
\citet{Eco}, for example,
  measures Pareto curves for virtual machines
  and combines them to achieve the best possible trade-off
  between memory and garbage collection time.
This data-driven approach works well for virtual machines,
  where long run times provide lots of data.
In a browser, where programs have a short lifetime,
  a model-based approach like MemBalancer is required.

More broadly, heap limit rules are merely
  one instance of the ever-present space-time trade-off.
This paper can be seen as automatically tuning
  one class of such trade-offs, but there are many others.
GCCache, for example, balances
  a Java application's software cache
  against its garbage collector~\cite{GCCache}.
The M3 system proposes a mechanism
  similar to V8's Memory Pressure Notifications
  to coordinate cache sizing across multiple applications~\cite{M3}.
Similar trade-offs are common throughout
  garbage-collected language runtimes;
  for example, in a jitted runtime,
  optimized code generation affects
  both runtime and memory use~\cite{Lite}.
Rematerialization, a compiler technique
  for reducing memory pressure,
  is a similar challenge, done statically~\cite{Remat}.
\citet{DTR} does something similar, though dynamically,
  for machine learning workloads.

\section{Conclusion}
This paper proposes
  that garbage-collected language runtimes
  should use a compositional heap limit rule
  to guarantee that multiple heaps
  allocate memory amongst themselves
  in a way that minimizes garbage collection time.
Standard heap limit rules, unfortunately,
  are not compositional,
  so we derive a compositional
  ``square-root'' limit rule,
  describe an algorithm that instantiates it,
  and implement that algorithm for V8
  in a system called MemBalancer.
On both memory-intensive benchmarks
  and real-world websites,
  MemBalancer leads to a significantly better
  trade-off between memory usage
  and garbage collection time.
On real world websites,
  reductions in average heap memory usage
  of up to \BROWSERIIMemorySaving are possible
  without increasing garbage collection time.

In the future, we hope
  to evaluate compositional heap limit rules
  for other virtual machines.
The other major JavaScript engines,
  Spidermonkey (used in Firefox)
  and JavaScriptCore (used in Safari),
  are natural fits,
  since both involve have multiple heaps
  when the user opens multiple tabs.
Investigating MemBalacer's impacts
  on server-side JavaScript frameworks
  such as node.js and Deno
  is also important;
  it's unclear if server-side applications
  show larger or smaller gains than web applications.
Compositional heap limit rules
  are also applicable to other languages,
  as demonstrated by the deployed Racket implementation.
(Though, as in the Racket implementation,
  language-specific simplifications may be appropriate.)
We are excited for experiments on other common language runtimes,
  such as in JVMs.
Most intriguingly,
  we hope to test whether
  com\-po\-si\-tio\-nal heap limit rules allow coordinating
  virtual machines for different languages,
  when for example a Racket program
  runs simultaneously with a web browser.

Theoretical improvements could also refine
  the notion of compositional heap limit rule.
Adding a model of memory fragmentation
  could allow better memory management
  in low-memory situations such as cheap smartphones.
A model of multiple generations
  could allow optimally sizing
  both the young and old generations,
  which could lead to further speed-ups.
Empirical work such as \cite{PerfGC}
  suggests that minor garbage collections
  have a different analytical model of run time
  so different heap limit rules would be compositional.
A model of garbage collection latency
  could reduce pause times
  and determine when to schedule
  concurrent marking phases
  such as those in V8.
A model of heap priorities
  could capture real-world valuations
  such as background tabs being less important
  than foreground tabs in a web browser.
Finally, automatically adjusting $c$
  to account for system memory usage,
  latency requirements,
  or other criteria
  could potentially make MemBalancer
  even more effective.

\begin{acks}
We thank the members of the V8 Garbage Collection team,
  including Hannes Payer, Michael Lippautz,
  Ulan Degenbaev, and Chris Hamilton,
  for explaining the architecture
  of the V8 garbage collector
  and suggesting implementation approaches.
We also thank
  the Firefox Spidermonkey team
  (including Steve Fink and Jon Coppeard)
  for reading and commenting on the paper,
  as well as
  Tucker Hermans, Ryan Stutsman, and Matthew Flatt
  for technical help.
Finally, we thank the anonymous reviewers
  for their comments and suggestions.
This work was funded by the National Science Foundation
  under Grant No. \grantnum{NSF}{2119939}
  and by a V8 Research Grant.

\end{acks}

\bibliography{main}

\end{document}